\documentstyle[prl,aps,twocolumn]{revtex}

\input{psfig.sty}
\begin{document}
\draft \font\Bbb =msbm10  scaled \magstephalf \def\id{{\hbox{\Bbb
I}}} \newcommand{\ket}[1]{\big| \, #1 \big\rangle}
\newcommand{\bra}[1]{\left \langle #1 \, \right |}
\newcommand{\proj}[1]{\ket{#1}\bra{#1}}
\newcommand{\braket}[2]{\langle\, #1\,|\,#2\,\rangle} \newcommand{\half}{\mbox{$\textstyle \frac{1}{2}$}} \def\opone{\leavevmode\hbox{\small1\kern-3.8pt\normalsize1}}
\newcommand{\tr}[1]{\mbox{Tr} \, #1 }
\def\emph#1{{\it #1}}
\def\textbf#1{{\bf #1}}
\def\textrm#1{{\rm #1}}

\title{Direct detection of quantum entanglement}

\author{Pawe{\l} Horodecki}
\address{Faculty of Applied Physics and Mathematics, Technical University of Gda\'nsk\\  80-952 Gda\'nsk, Poland.}

\author{Artur Ekert} \address{Centre for Quantum Computation, Clarendon Laboratory, \\  The University of Oxford, Parks Road, Oxford OX1 3PU, UK.}

\maketitle

\begin{abstract}
\textbf{ Quantum entanglement, after playing a significant role in
the development of the foundations of quantum
mechanics~\cite{Schr35,EPR35,Bell64}, has been recently
rediscovered as a new physical resource with potential commercial
applications such as, for example, quantum
cryptography~\cite{Eke91}, better frequency
standards~\cite{BIWH96} or quantum-enhanced positioning and clock
synchronization~\cite{GLM01}. On the mathematical side the studies
of entanglement have revealed very interesting connections with
the theory of positive maps~\cite{HHH96,ABH3RWWZ}. The capacity to
generate entangled states is one of the basic requirements for
building quantum computers. Hence, efficient experimental methods
for detection, verification and estimation of quantum entanglement
are of  great practical importance. Here, we propose an
experimentally viable, \emph{direct} detection of quantum
entanglement which is efficient and does not require any \emph{a
priori} knowledge about the quantum state. In a particular case of
two entangled qubits it provides an estimation of the amount of
entanglement. We view this method as a new form of quantum
computation, namely, as a decision problem with quantum data
structure.} \end{abstract}

\bigskip

Suppose we are given $n$ pairs of particles, all in the same
quantum state described by some density operator $\varrho$, which
is unknown. We need to decide whether the particles in each pair
are entangled or not. From a mathematical point of view we need to
assert whether $\varrho$ can be written as a convex sum of product
states~\cite{Wer89}, \begin{equation} \varrho=\sum_{i}^{k} p_i\;
|\alpha_i\rangle\langle\alpha_i| \otimes
\beta_i\rangle\langle\beta_i|, |\label{sep}
\end{equation}
with $\ket{\alpha_i}$ and $\ket{\beta_i}$ pertaining to different
particles in the pair, and $\sum_i p_i=1$. It is assumed that the
Hilbert spaces associated with each particle are of finite
dimensions $d$ (taken to be the same for the two particles), so
that one can always find $k\le d^2$. If $\varrho$ were known then
we could try either to find the decomposition (\ref{sep}) directly
or to use one of the mathematical separability
criteria~\cite{ABH3RWWZ}. For sufficiently large $n$ we may indeed
start with the quantum state estimation, however, this involves
estimating $d^4-1$ real parameters of $\varrho$, most of which are
irrelevant in the context of the entanglement detection. In the
following we describe a direct method of detecting quantum
entanglement without invoking the state estimation.

We construct a measurement which can be performed on all copies of
$\varrho$ and which is as powerful in detecting quantum
entanglement as the best mathematical test based on positive
maps~\cite{HHH96}. The measurement can be viewed as two
consecutive physical operations: firstly, we construct a
transformation which maps $\varrho$ into an appropriate state
$\varrho'$ and, secondly, we measure the lowest eigenvalue of
$\varrho'$. This eigenvalue alone serves as a separability
indicator.

A convenient starting point for our construction is the most
powerful, albeit purely mathematical and not directly
implementable, separability criterion proposed to date. It is
based on mathematical properties of linear positive maps acting on
matrices~\cite{HHH96}. Let $M_d$ be a space of matrices of
dimension $d$; recall that $\Lambda : M_d \mapsto M_d$ is called
positive if $X\ge 0$ implies $\Lambda (X)\ge 0$ (expression $X\ge
0$ means that the matrix $X$ has a nonnegative spectrum). If the
induced map $\id \otimes \Lambda$ is also positive then $\Lambda$
is called completely positive, and, as such, it represents a
physically allowed transformation of density operators (here $\id$
denotes the identity map on an auxiliary system of any dimension).
Using this terminology the separability criterion reads
~\cite{HHH96}: $\varrho$ is separable iff

\begin{equation}
[\id \otimes \Lambda](\varrho)\geq 0, \label{maps}
\end{equation}
for all positive but not completely positive maps $\Lambda: M_d
\mapsto M_d$ acting on the second particle. In fact it is
sufficient to consider only positive maps $\Lambda$ such that the
maximum of $\tr\Lambda(\varrho)$ over all $\varrho$ is equal to
unity. Other positive maps differ only by a positive
multiplicative factor which does not affect the
condition~(\ref{maps}).

Furthermore, in some cases, instead of scanning all positive maps,
we can choose just one. For example, \emph{all} entangled states
of two qubits can be detected by choosing $\Lambda$ to be
transposition~\cite{Per96,HHH96}. The snag is that positive maps
$\Lambda$, such as an anti-unitary transposition, and the induced
maps $\id\otimes\Lambda$ cannot be implemented in a laboratory.
Thus, the criterion~(\ref{maps}) tacitly assumes prior knowledge
of $\varrho$. However, there is a way to modify it so that it
becomes experimentally viable without involving any state
estimation.

If we mix in an appropriate proportion $[\id\otimes\Lambda]$ with
a depolarizing map that turns any density matrix into a maximally
mixed state then the resulting map can be completely positive This
is because the lowest negative eigenvalues generated by the
induced map $[(\id\otimes\id)\otimes(\id\otimes\Lambda)]$ can be
offset by the positive eigenvalues of the maximally mixed state
generated by the depolarizing map. The most negative eigenvalue
$-\lambda<0$ is obtained when
$[(\id\otimes\id)\otimes(\id\otimes\Lambda)]$ acts on the
maximally entangled state of the form
$\frac{1}{\sqrt{d^2}}\sum_{i=1}^{d^2} |i\rangle|i \rangle $, where
each state $\ket{i}$ pertains to a $d^2$ dimensional subsystem
which itself is composed of two $d$ dimensional parts. Thus the
map

\begin{equation}
[\widetilde{\id\otimes\Lambda}](\varrho)= p \frac{I \otimes
I}{d^{2}} + (1-p) [\id\otimes\Lambda](\varrho),
\label{fizycznamapa} \end{equation} is completely positive and
therefore physically implementable when the induced map
$[(\id\otimes\id)\otimes(\widetilde{\id\otimes\Lambda)}]$ is
positive, which happens for $p \ge (d^{4}\lambda)/(d^{4} \lambda +
1)$ ~\cite{nonlinear}. By inserting the threshold value
$p=(d^{4}\lambda)/(d^{4} \lambda + 1)$ into (\ref{fizycznamapa})
we can modify the criterion~(\ref{maps}) as follows: $\varrho$ is
separable iff for all positive maps $\Lambda$,

\begin{equation}
[\widetilde{\id\otimes\Lambda}](\varrho)\ge
\frac{d^{2}\lambda}{d^{4} \lambda +1}, \label{relation}
\end{equation} i.e. when the minimal eigenvalue of the transformed
state $\varrho'=[\widetilde{\id\otimes\Lambda}](\varrho)$ is
greater than $(d^{2}\lambda)/(d^{4} \lambda +1)$. In general, for
some maps $\Lambda$, the related completely positive maps
$\widetilde{\id \otimes \Lambda}$ are not trace-preserving and
require postselections in their physical implementations. Maps
such as $\widetilde{\id \otimes \Lambda}$ have been referred to as
``structural'' physical approximations of unphysical maps $\id
\otimes \Lambda$~\cite{nonlinear}.

For example, if we take $\Lambda$ to be transposition $T$, (the
first positive map used for detecting entanglement), we obtain
\begin{equation} [\widetilde{\id\otimes T}](\varrho) =
\frac{d}{d^3 +1} I\otimes I + \frac{1}{d^3 +1} [\id \otimes
T](\varrho). \label{transposition} \end{equation} In the two qubit
case, where the partial transposition is a sharp test for
entanglement, we obtain, \begin{equation} [\widetilde{\id\otimes
T}](\varrho)= \frac{2}{9} I \otimes I + \frac{1}{9} [\id \otimes
T](\varrho), \end{equation} which can be represented and
implemented as

\begin{equation} \frac{1}{3}
\Lambda_{1}\otimes \Lambda_{2} +\frac{2}{3}\id \otimes
\sigma_{x}\sigma_{z}\Lambda_{1} \sigma_{z} \sigma_{x}, \label{2q}
\end{equation} with the two channels defined as:

\begin{equation}
\Lambda_{1}(\varrho)=\frac{1}{3} \sum_{i=x,y,z} \sigma_{i} \varrho
\sigma_{i}, \quad \Lambda_{2}(\varrho)=
\frac{1}{4}\sum_{i=o,x,y,z} \sigma_{i} \varrho \sigma_{i}.
\end{equation}

The map can be implemented by applying selected products of
unitary (Pauli) transformations with the prescribed probabilities.
The map  It is trace-preserving hence any postselection in
experimental realizations is avoided.

Thus, in order to detect entanglement of an arbitrary two-qubit
state $\varrho$ it is enough to estimate a single parameter, i.e.
the minimal eigenvalue of $[\widetilde{\id\otimes T}](\varrho)$.
The state $\varrho$ is separable iff this eigenvalue satisfies
$\lambda_{min}\geq \frac{2}{9}$. Let us also point out an extra
bonus: $\lambda_{min}$ gives us $-\lambda'$, the most negative
eigenvalue of $[\id\otimes T](\varrho)$, which enters the
expression for the upper and lower bounds for the entanglement of
formation,

\begin{eqnarray}
&&H\left(\frac{1+\sqrt{1-4\lambda'^2}}{2}\right) \leq E(\varrho)
\nonumber \\ &&\leq
H\left(\frac{1+\sqrt{1-4\left(\sqrt{2\lambda'^{2}+\lambda'}-\lambda'\right)}}{2}\right),
\end{eqnarray}
where $H(x)$ is the Shannon entropy. The above formulae can be
derived from the estimations of the concurrence provided by
Verstraete et al~\cite{VADM01}.

Suppose for a moment that $\widetilde{\id \otimes \Lambda}$ is
trace-preserving, e.g. the transposition case. The first part of
our entanglement detection measurement is accomplished by applying
$\widetilde{\id\otimes\Lambda}$ to each of the $n$ pairs to obtain
$n$ copies of $\varrho'=[\widetilde{\id\otimes\Lambda}](\varrho)$.
Then, following the criterion~(\ref{relation}), we need to measure
the lowest eigenvalue of $\varrho'$.

This can be viewed as a special case of the spectrum estimation
and possible approaches depend a lot on particular physical
realizations of $\varrho'$. Here, we provide two general
solutions. The first one, based on quantum interferometry, is
conceptually simple and relies on estimating $d^2-1$ parameters
from which the spectrum of $\varrho'$ can be calculated (this is a
significant gain over the state estimation which involves $d^4-1$
parameters). The second solution is a joint measurement on all
copies of $\varrho'$ which gives directly the estimate of the
lowest eigenvalue.

\begin{figure}[ht]
\vskip .1cm
\centerline{\psfig{file=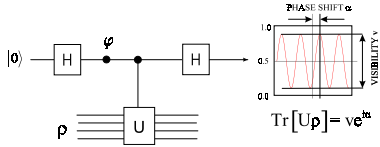,width=70mm}} \vskip
.4cm \caption{Both the visibility and the shift of the
interference patterns of a single qubit (top line) are affected by
the controlled-$U$ operation. This set-up allows to estimate $\tr
U\rho$, the average value of  $U$ in state $\rho$.} \label{figint}
\end{figure}

We start with the quantum interferometry, presented here as a
quantum network shown in Fig.(\ref{figint}). A typical
interferometric set-up for a single qubit --- the Hadamard gate,
phase shift $\phi$, the Hadamard gate, followed by a measurement
--- is modified by inserting in between the Hadamard gates a
controlled-$U$ operation, with its control on the qubit and with
$U$ acting on a quantum system described by some unknown density
operator $\rho$. (N.B. we do not assume anything about the form of
$\rho$, it can, for example, describe several entangled or
separable sub-systems.) The action of the controlled-$U$ on $\rho$
modifies the interference pattern by the factor,

\begin{equation}
{\rm Tr}\rho U = {\rm v} e^{i\alpha}, \label{eqvisi}
\end{equation} where $\rm v$ is the new visibility and $\alpha$ is
the shift of the interference fringes, also known as the
Pancharatnam phase~\cite{Pancha56}. Formula~(\ref{eqvisi}) has
been derived, in the context of geometric phases, by Sj\"{o}qvist
et al.~\cite{SPEAEOV00}.

The network can evaluate certain non-linear functionals of density
operators. Indeed, let us choose $\rho$ to be composed of two
subsystems, $\rho=\varrho_a\otimes\varrho_b$, and let $U$ to be
the exchange operator $V$ such that $V\ket{\alpha}\ket{\beta}
=\ket{\beta}\ket{\alpha}$ for any pure states of the two
subsystems. The interference pattern is now modified by the factor
$\tr\, V(\varrho_a\otimes\varrho_b) = \tr\,\varrho_a\varrho_b$.
For $\rho=\varrho\otimes\varrho$ we can estimate $\tr \varrho^2$,
which gives us an estimate of $\sum_{i=1}^m \lambda_i^2$, where
$\lambda_i$ are the eigenvalues of $\varrho$. N.B. $\tr\varrho^2$
is real hence there is no need to sweep the phase $\varphi$ in the
interferometer, it can be fixed at $\varphi=0$.

In general, in order to calculate the spectrum of any $m \times m$
density matrix $\varrho$ we need to estimate $m-1$ parameters
${\rm Tr}\varrho^2$, ${\rm Tr}\varrho^3$,... ${\rm Tr}\varrho^m$.
For this we need the controlled-shift operation. Given $k$ systems
of dimension $m$ we define the shift $V^{(k)}$ as

\begin{equation}
V^{(k)} \ket{\phi_1}\ket{\phi_2}...\ket{\phi_k} =
\ket{\phi_k}\ket{\phi_1}...\ket{\phi_{k-1}},
\end{equation}
for any states $\ket{\phi}$. Such an operation can be easily
constructed by cascading $k-1$ swaps $V$. This time, if we prepare
$\rho=\varrho^{\otimes k}$ the interference will be modified by
the factor

\begin{equation}
{\rm Tr}\, \varrho^{\otimes k}V^{(k)} = {\rm Tr}\, \varrho^k =
\sum_{i=1}^m \lambda_i^k. \label{powers}
\end{equation} Thus
measuring the average values of $V^{(k)}$ for $k=2,3...m$ gives us
effectively the spectrum of $\varrho$. In particular, in our case,
we obtain the spectrum (and the lowest eigenvalue) of
$\varrho'=[\widetilde{\id \otimes \Lambda}](\varrho)$ by
estimating $d^2-1$ parameters $\tr \varrho'^k$, where $k=2...
d^2$. Again, the phase in the interferometry can be fixed at
$\varphi=0$.

The interferometric scheme described above is conceptually simple
and experimentally viable, however, if the simplicity of the
implementation is not an issue then we can measure the estimate of
the lowest eigenvalue directly. This requires a join measurement
on all of the $n$ pairs. We use the Keyl and Werner spectrum
estimation method~\cite{KW01}, which, in the entanglement
detection context, works as follows. The $n$ copies of the
$m\times m$ state $\varrho'$ (in our case $m=d^2$) form an
operator on the n-fold tensor product space which can be
decomposed according to irreps of $SU(m)$, so that each summand,
including multiplicities, is labelled by a Young tableau, i.e. $n$
boxes arranged in rows of decreasing length (c.f.~\cite{CEM99} for
the $SU(2)$ case). The tableaus give a family of projectors for
the spectrum estimation measurement. The normalized row lengths of
each tableau are taken as estimates of the ordered sequence of
eigenvalues of $\varrho'$. The probability that the error is
greater than some fixed $\epsilon$ decreases {\it exponentially}
with $n$~\cite{KW01}. In our particular case, we are interested
only in the lowest eigenvalue. We modify the Keyl-Werner scheme by
adding together all projectors corresponding to Young tableaus
with the fixed length of the last row. The measurement determined
by these projectors gives directly the estimate of only one
parameter --- the lowest eigenvalue of $\varrho'$. Such a
measurement can be represented as a quantum network implementing
projections on the symmetric and on partially symmetric subspaces
(see~\cite{BDEJM97} for the network projecting on the symmetric
subspace).

Our considerations remain valid, with some minor modifications,
when $\widetilde{\id \otimes \Lambda}$ is not trace-preserving. In
this case experimental implementations require postselections,
which result in $n'=n\tr(\widetilde{\id \otimes \Lambda}(\varrho)$
copies of normalized states $\widetilde{\id \otimes
\Lambda}(\varrho) /\tr(\widetilde{\id \otimes \Lambda}(\varrho))$.
The spectrum estimation procedure is not affected, however, before
checking the condition~(\ref{relation}) the lowest eigenvalue has
to be rescaled by  the factor $\tr(\widetilde{\id \otimes
\Lambda}(\varrho))$.

Let us summarize our findings. Given $n$ copies of a bipartite
$d\otimes d$ system described by some unknown density operator
$\varrho$ we can test for entanglement either by estimating
$\varrho$ and applying criterion~(\ref{maps}), or, more directly,
by performing the measurements we have just described. The state
estimation involves estimating $d^4-1$ parameters of $\varrho$,
most of which are of no relevance for the entanglement detection.
The optimal state estimations rely on joint measurements on all
copies of $\varrho$. However, one can construct less efficient but
simpler state estimation methods which involve measurements only
on individual copies. Our more direct, interferometry based,
method requires estimations of only $d^2-1$ parameters and joint
operations on $d$ copies of $\varrho'$. The most demanding, from
the experimental point of view, is our second method. It is a
measurement with an outcome which is an estimate of just one
parameter, but, like the optimal state estimation, the measurement
involves joint operations on all copies of $\varrho'$. Both direct
and indirect entanglement detections have their own merits.
Depending on the context, applications, and technologies involved
one can choose one or the other.

Direct entanglement detections, can be employed as sub-routines in
quantum computation. For example, one may consider performing or
not performing a quantum operation on a given quantum system
conditioned on some part of quantum data being entangled or not.
In fact direct entanglement detections can be viewed as quantum
computations solving an {\it inherently quantum} decision problem:
given as an input $n$ copies of $\varrho$ decide whether $\varrho$
is entangled. Here the input data is quantum and such a decision
problem cannot even be even formulated for classical computers.
Nonetheless the problem is perfectly well defined for quantum
computers. Finally, let us add that the method presented here can
be easily generalized to cover all linear maps tests for {\it
arbitrary} multiparticle entanglement~\cite{Ho00PLA} and the so
called k-positive map tests detecting Schmidt numbers of density
matrices \cite{Schmidt}.
Modification of the method to the case of two distant labs
(i. e. under restrictions to local operations and classical communication
\cite{ABH3RWWZ}) will be considered elsewhere.

For the sake of completeness we should also mention here that
there are two-particle observables, called entanglement witnesses
which can detect quantum entanglement is some special cases (see
\cite{witnesses,ABH3RWWZ}). They have positive mean values on {\it
all} separable states and negative on {\it some} entangled states.
Therefore any individual entanglement witness leaves many
entangled states undetected. When $\varrho$ is unknown we need to
check infinitely many witnesses, which effectively reduces this
approach to the quantum state estimation. However, let us point
out any witness defines a positive map which can be used in our
test.

To conclude, we have demonstrated that direct and physically
implementable methods of entanglement detection are possible. They
are equivalent to the most powerful mathematical separability
criteria known to date~\cite{HHH96}. We have described their
possible realizations in the generic terms of quantum gates and
networks. These generic components can be implemented using
several experimental techniques ranging from trapped ions to
quantum dots~\cite{BEZ}.

This research was supported in part by the Polish Committee for
Scientific Research, European Commission, Elsag Spa, EPSRC, and
the Royal Society, London.


\end{document}